\documentclass[twocolumn,aps,pra,showpacs,groupedaddress]{revtex4-1}
\usepackage{graphicx,amssymb,amsmath,mathrsfs}
\usepackage[matrix,frame,arrow]{xy}%    Q-circuit version 1.2
\usepackage[matrix,frame,arrow]{xy}
\usepackage{amsmath}

\newcommand{\qw}[1][-1]{\ar @{-} [0,#1]}
\newcommand{\multigate}[2]{*+<1em,.9em>{\hphantom{#2}} \qw \POS[0,0].[#1,0];p !C *{#2},p \save+LU;+RU **\dir{-}\restore\save+RU;+RD **\dir{-}\restore\save+RD;+LD **\dir{-}\restore\save+LD;+LU **\dir{-}\restore}
\newcommand{\ghost}[1]{*+<1em,.9em>{\hphantom{#1}} \qw}
\newcommand{\ustick}[1]{*!D!<0em,-.5em>=<0em>{#1}}
\newcommand{\Qcircuit}[1][0em]{\xymatrix @*=<#1>}
\newcommand{\pureghost}[1]{*+<1em,.9em>{\hphantom{#1}}}
\newtheorem{lemma}{Lemma} 

\newtheorem{corollary}{Corollary} 
\newtheorem{theorem}{Theorem}
 
\newtheorem{definition}{Definition}
\def\spc #1{{\mathsf #1}} 
\def\Proof{\medskip\par\noindent{\bf Proof.  }} 
\def\qed{$\,\blacksquare$\par} \def\>{\rangle} 
\def\<{\langle}
 \def\map#1{\mathscr #1}
 \def\sys#1{\mathrm{#1}} \def\rA{{\sys A}}\def\rB{{\sys B}}
\def\rC{{\sys C}}
  
 \def\rI{{\sys I}}
  
\def\rX{\sys X}
\def\rZ{\sys Z}
\def\rem{\sys m}\def\rn{\sys n}
\def\rp{\sys p}
\def\samp#1{\mathrm{#1}}   
\def\sH{{\spc H}} 
\def\sL{{\spc L}} 
\def\tA{\map A}

\def\tI{\map I}\def\tT{\map T}

\def\tT{\map T} 
\def\Stset{\mathsf{St}}

\def\Stsep{\mathsf{St}_+}

\def\Span{\mathsf{Span}} 
\def\Trnset{{\mathsf T}} 
\def\Trnsep{{\mathsf T}_+}

\def\aff{\mathsf{Aff}}
\def\con{{\mathsf T}_+}
\def\set#1{\mathsf{#1}}

\def\herm{\mathsf {Herm}}
\def\det{\Trnset_1}\def\pos{\mathsf P}\def\Herm{\mathsf {Herm}}

\def\Tr{\mathrm{Tr}}
   \def\Reals{{\mathbb R}}
  \def\ch{\ensuremath{\mathsf{Ch}}} 
  
  \def\bvec#1{\mathbf{#1}}
  \def\bi{\bvec i}
    \def\vb{\bvec b}
\begin{document}

\title{Causal structures and the classification of higher order quantum computations}
\author{Paolo Perinotti}\email{paolo.perinotti@unipv.it} 
\affiliation{{\em QUIT Group}, Dipartimento di Fisica  and INFN, via Bassi 6, 27100 Pavia, Italy}
\homepage{http://www.qubit.it}
\date{\today}
\begin{abstract}
\end{abstract}
\begin{abstract}
Quantum operations are the most widely used tool in the theory of quantum information processing, representing elementary transformations of quantum states that are composed to form complex quantum circuits. The class of quantum transformations can be extended by including transformations on quantum operations, and transformations thereof, and so on up to the construction of a potentially infinite hierarchy of transformations. In the last decade, a sub-hierarchy, known as quantum combs, was exhaustively studied, and characterised as the most general class of transformations that can be achieved by quantum circuits with open slots hosting variable input elements, to form a complete output quantum circuit. The theory of quantum combs proved to be successful for the optimisation of information processing tasks otherwise untreatable. In more recent years the study of maps from combs to combs has increased, thanks to interesting examples showing how this next order of maps requires entanglement of the causal order of operations with the state of a control quantum system, or, even more radically, superpositions of alternate causal orderings. Some of these non-circuital transformations are known to be achievable and have even been achieved experimentally, and were proved to provide some computational advantage in various information-processing tasks with respect to quantum combs. Here we provide a formal language to form all possible types of transformations, and use it to prove general structure theorems for transformations in the hierarchy. We then provide a mathematical characterisation of the set of maps from combs to combs, hinting at a route for the complete characterisation of maps in the hierarchy. The classification is strictly related to the way in which the maps manipulate the causal structure of input circuits.
\end{abstract}

\maketitle

\section{Introduction}
The explosion of the field of quantum information theory \cite{nielchu}, and quantum computation in particular, is largely based on the framework of {\em quantum circuits} \cite{deutsch,yao}, that provides an abstract language for the representation of quantum algorithms---sequences of quantum operations performed in a precise order on a given input state. The building blocks of quantum circuits are {\em quantum gates}, elementary unitary operations on one or more qubits, along with very special operations corresponding to the preparation of a reset state or measurement in the so-called computational basis.

While standard quantum circuits evolve pure quantum states unitarily, this language can be generalised to encompass evolution of mixed states via irreversible channels \cite{irrevcirc}. Thus, in the generalised framework the primary notion becomes that of a {\em quantum instrument}, a collection of transformations labeled by an {\em outcome}---the value of a classical variable---representing a conditional evolution within a chosen test. The quantum instrument provides the description of what is generally referred to as {\em state reduction} after a quantum measurement. 

Quantum circuits are then the language for description of input-output flow of information in the processing of a quantum state. The classical counterpart of such a processing is a function (here we consider general, possibly irreversible computations, and thus the function can be more generally a probabilistic map) that transforms input bit strings to output strings. One normally identifies the abstract input-output flow in a circuit with the time evolution of the corresponding systems implementing the algorithm. The identification of time evolution with the input-output direction is a consequence of {\em causality} \cite{purif,deriv}, the property of quantum theory (and of classical information theory as well) that forbids communication from the output towards the input.

What is peculiar about quantum channels and instruments is that one can define them axiomatically---as maps on states that must only comply to the requirement of providing positive and normalised probability distributions when used in a closed circuit. No further requirement is necessary to identify {\em physical} transformations, since all the conceivable quantum instruments satisfy a realisation theorem in terms of standard unitary evolutions and projective quantum measurements \cite{stine,kraus,ozawa}, granting that at least in principle they all correspond to implementable processes. 

What happens if we now consider abstract maps from quantum channels to quantum channels, or from quantum instruments to quantum instruments? Is it sufficient for such a map to respect the properties of probabilities to be feasible in practice? And what if we continue constructing higher and higher orders of maps? What is known so far is that for a sub-hierarchy of maps---called quantum combs \cite{qca,comblong} and encompassing all conceivable strategies in a quantum game \cite{watgut}---compatibility with probability theory is sufficient for feasibility.

%%%%%%%%%%%%%%%%%%%%%%%% Check the statement here! %%%%%%%%%%%%%%%%%%%

However, the construction of a mathematical hierarchy of functions on functions can continue arbitrarily far. In this article we introduce a rigorous language that allows us to deal with the full hierarchy of transformations, and we use it to provide the first classification result for all maps respecting the basic probabilistic structures. As it was noticed in Ref.~\cite{switch}, there exist maps in the hierarchy that call for a generalisation of the quantum circuit framework, being operationally feasible, but not circuital. Moreover, there are conceivable maps that do not have any interpretation in terms of presently known physical schemes \cite{orecobru}, and could be conceivable provided that the causal ordering of operations could be entangled with the state of a quantum control system. 

%%%%%%%%%%%%%%%%%%%%%%%%%%%%%%%%%%%%%%%%%%%%%%%%%%%%%%%%%%

Interestingly, the existence of admissible maps that cannot be reduced to a definite causal structure is proved also in the case of classical circuits \cite{wolfbau}.

Quantum gates constitute the first-order of the hierarchy of transformations, and the full hierarchy of higher-order maps is then based on a causal theory, where there is a notion of {\em computational time}, whose connection with physical time is straightforward. The causal structures of first-order maps provide a sort of imprint on the full hierarchy, imposing constraints for the definition of higher-order maps at all levels. 

In section \ref{s:prelim} we introduce the basic mathematical definitions and list the main results in the literature about first-order maps. In section \ref{s:hier} we introduce typing rules, that allow us to construct and express all conceivable types of higher-order maps, and prove our main classification result. In section \ref{s:conc} we conclude with remarks and comments about the main open questions.

\section{Mathematical preliminaries}\label{s:prelim}

The theory of quantum computation deals with transformations of quantum systems. A quantum system can be the spin of a particle, the polarization of a photon, the current of a superconducting circuit, etc. We will denote systems by roman capital letters $\rA$, $\rB,\ldots,\rZ$. In abstract terms, the characterising property of a quantum system is the number of its effective degrees of freedom. Thus, when we refer to a system $\rA$ we mean any physical system with a given dimension $d_\rA$. A special role is played by the system with dimension 1, the {\em trivial system}, that will be denoted by $\rI$. Quantum systems are in correspondence with complex Hilbert spaces, thus for a system $\rA$ we will have $\sH_\rA\simeq\mathbb C^{d_\rA}$. The {\em parallel composition} of systems $\rA$ and $\rB$ is denoted by $\rA\rB$ and is the system $\rC$ with $d_\rC=d_\rA d_\rB$.

The set $\Stset(\rA)$ of {\em states} of system $\rA$ is the set of sub-normalised density matrices on $\sH_\rA$, non-negative-definite operators $\rho\geq0$ on $\sH_\rA$ with $\Tr[\rho]\leq1$. The trace of a state represents its preparation probability. The set $\Stset_1(\rA)$ of deterministic states is the set of those states $\rho$ with $\Tr[\rho]=1$. The real span of density matrices on $\sH_\rA$ is the space of Hermitian operators $\Herm(\rA)$, that we will denote by $\Stset_{\mathbb R}(\rA)$.

A linear map $\tA\in\Trnset(\rA\to\rB)$ is {\em completely positive} if $\tA\otimes\tI_\rC(\rho)\geq0$ for every $\rho\in\Stset(\rA\rC)$ and for every system $\rC$. The map $\tA$ is {\em trace non-increasing} if $\Tr[\tA(\rho)]\leq\Tr[\rho]$ for every $\rho\in\Stset(\rA)$. The set of {\em quantum operations} $\Trnset(\rA\to\rB)$ corresponds to completely positive trace non-increasing maps from operators on $\sH_\rA$ to operators on $\sH_\rB$.  Deterministic transformations correspond to {\em trace-preserving} maps, satisfying $\Tr[\tA(\rho)]=\Tr[\rho]$ for every state $\rho\in\Stset(\rA)$. The set of deterministic transformations $\rA\to\rB$ is denoted by $\det(\rA\to\rB)$, and its elements are called {\em channels}.

A {\em quantum instrument} from system $\rA$ to system $\rB$ is a family $\{\tA_i\}_{i\in\rX}\subseteq\Trnset(\rA\to\rB)$ such that $\tA:=\sum_{i\in\rX}\tA_i\in\det(\rA\to\rB)$. Since linear maps can be linearly combined, we construct the real space $\Trnset_{\mathbb R}(\rA\to\rB)$ by linear extension of $\Trnset(\rA\to\rB)$. The cone of completely positive maps will be denoted by $\Trnset_+(\rA\to\rB)$.

A very useful way to represent transformations is through the Choi-Jamio\l kowski isomorphism \cite{choi}, a linear mapping from the space $\Trnset_{\mathbb R}(\rA\to\rB)$ to $\Stset_{\mathbb R}(\rA\rB)$. The mapping is defined as
\begin{align}
&\ch(\tA):=\tA\otimes\tI_{\rA'}(|\Omega\>\<\Omega|),\nonumber\\
&[\ch^{-1}(R)](\rho):=\Tr_{\rA}[(I_\rB\otimes\rho^T)R],
\end{align}
where $d_{\rA'}=d_\rA$ and $|\Omega\>\in\sH_{\rA\rA'}$ is the vector $|\Omega\>:=\sum_{n=1}^{d_\rA}|n\>_{\rA}|n\>_{\rA'}$, $\{|n\>\}_{n=1}^{d_\rA}$ denoting a choice of canonical orthonormal bases in $\sH_\rA$ and $\sH_{\rA'}$, and $X^T$ denotes transposition of the operator $X$ in the canonical basis. The main reason of interest in the Choi-Jamio\l kowski isomorphism is that it provides a necessary and sufficient condition for complete positivity as follows
\begin{align}
\tA\in\Trnsep(\rA\to\rB)\ \Leftrightarrow \ch(\tA)\in\Stsep(\rB\rA).
\end{align}
The trace non-increasing completely positive maps are precisely those whose Choi-Jamio\l kowski image satisfies
\begin{align}
\Tr_\rB[\ch(\tA)]\leq I_\rA,
\label{eq:norma}
\end{align}
with trace-preserving maps saturating the inequality.

A special type of instrument is given by POVMs, which transform states into probabilities, namely transformations $\rA\to\rI$. A POVM for system $\rA$ is thus a collection of positive operators on $\sH_\rA$ that sums to the identity, as from Eq.~\eqref{eq:norma} with $d_\rB=1$. Notice also that states of system $\rB$ can be considered as a special case of completely positive maps from $\rI$ to $\rB$, where the (sub-)normalisation constraint is simply given by Eq.~\eqref{eq:norma} for $d_\rA=1$.

The picture of quantum states, quantum operations and effects provides the complete description of quantum circuits, which correspond to processes obtained by an ordered composition of elementary instruments.

\section{The hierarchy}\label{s:hier}

We will now introduce higher order computation, by enlarging the class of transformations that we consider. In particular, this is obtained by enriching the way in which we can compose systems to get new systems. The new composition rule was implicitly already used when we introduced transformations, to which we attributed a type $\rA\to\rB$. However, since now we want to use objects of type $\rA\to\rB$ as inputs and outputs of a new type of transformations, we need to define the construction of new types thoroughly. This is achieved by the following recursive definition.

Given two types $x,y$, one can form the type $x\to y$. In particular, as a shorthand notation, $\overline x:=x\to \rI$ will denote positive linear functionals bounded by 1 on elements of type $x$. We also formally define a new composition law $\otimes$ of types as follows. For every couple $x,y$, 
\begin{align*}
x\otimes y:=\overline{x\to \overline y}.
\end{align*}

\begin{definition}
A {\em deterministic event of type $x\to y$} is the Choi representative of an admissible transformation from events of type $x$ to events of type $y$ such that the image of every deterministic event of type $x$ is a deterministic event of type $y$. An {\em event of type $x\to y$} is a positive operator $S$ such that $S\leq R$ for
  some deterministic event $R$ of type $x\to y$.
\end{definition}

As a remark, we stress that the above notion of {\em admissibility} means that if you have a map $\map M$ of type $x\to y$, and apply $\map M\otimes \map I_z$ on an event of type $x\otimes z$, what you obtain is an event of type $y\otimes z$, for arbitrary type $z$. We will prove in the following that this condition is equivalent to complete positivity for every type in the hierarchy. The consistence of this conclusion can be seen form the fact that the Choi representative of a completely positive map is a positive operator, and thus, loosely speaking, admissibility corresponds to the preservation of positivity under local application of the map. For the above reason, it is useful to introduce a symbol for the set of positive operators on $\sH_x$, i.e.
\begin{align*}
\pos(\sH_x):=\{X\in\Herm(\sH_x)|X\geq0\}.
\end{align*}

If we denote by $\sH_x$ the Hilbert space on which events of type $x$ are defined, we clearly have
\begin{align}
\sH_{x\to y}=\sH_y\otimes\sH_x.
\end{align}
Moreover, since $\sH_\rI=\mathbb C$, one has
\begin{align}
\sH_{\overline x}=\sH_x,
\end{align}
and finally
\begin{align}
\sH_{x\otimes y}=\sH_y\otimes\sH_x.
\end{align}
The convex set of deterministic events
$\Trnset_1(x)$ also determines the convex set of events of type $x$, denoted by $\Trnset(x)$, as the set of Choi representatives of admissible maps dominated by $Z\in\det(x)$. From this point of view the cone $\con(x):=\{K\in\herm(\sH_x)|\exists\lambda>0,R\in\Trnset(x):\ K=\lambda R\}$ is not sufficient to specify a type. As a trivial example, consider the cones  $\con(\rI\to\rA)$ and $\con(\rA\to\rI)$: They are the same, but the types $\rI\to\rA$ (states of $\rA$) and $\rA\to\rI$ (effects of $\rA$) are different because of very different normalisation constraints. We then
introduce the following identity criterion for types.
\begin{definition}\label{def:equitypes}
  We say that two types $x$ and $y$ are {\em equivalent}, and denote
  it as $x= y$, if $\con(x)=\con(y)$ and $\Trnset_1(x)=\Trnset_1(y)$.
\end{definition}

Given this definition, we can show that $\rA\otimes\rB$ is the parallel composition of systems $\rA\rB$
\begin{lemma}
The type $\rA\otimes\rB$ coincides with the parallel composition of systems $\rA\rB$.
\end{lemma}
\Proof Let us first determine the most general map $\rA\to\overline\rB$. Its Choi is a positive operator on $\sH_\rA\otimes\sH_\rB=\sH_{\rA\rB}$. A deterministic map of this kind corresponds to a positive operator $Q$ such that for all $\rho_\rB\geq0$ with $\Tr[\rho_\rB]=1$ one has
\begin{align*}
\Tr_\rB[Q(I\otimes\rho_\rB^T)]=\rI_\rA.
\end{align*}
This means that for every $\sigma_\rA\otimes \rho_\rB$ with $\Tr[\sigma]=\Tr[\rho]=1$ one has
\begin{align*}
\Tr[Q(\sigma\otimes\rho^T)]=1,
\end{align*}
and by the polarisation identity this implies $Q=I_{\rA\rB}$. Thus, events of type $\rA\to\overline\rB$ are positive operators bounded by $I$, namely they coincide with the set of effects of $\rA\rB$. Finally, positive functionals bounded by 1 on these events coincide with states of the system $\rA\rB$.
\qed

The construction of types through the composition rule ``$\to$" allows us to prove properties $P$ of types by induction, by proving it for every elementary type $A$, namely $P(A)=1$, and then proving that $P(x)=P(y)=1\Rightarrow P(x\to y)=1$.
As an example, we now prove two crucial lemmas.

\begin{lemma}\label{lem:firstrecpro}
The convex set $\det(x)$ of deterministic events of type $x$ is the set of all positive operators of the form 
\begin{align}
X=\lambda_xI_x+T,
\label{eq:decompdet}
\end{align}
where $\lambda_x$ is a suitable constant $\lambda_x>0$, and the operators $T$ span a suitable subspace $\Delta_x$ of the real space of traceless selfadjoint operators on $\sH_x$. In particular, the operator $\lambda_xI_x$ represents a deterministic event.
\end{lemma}

\Proof The thesis is true for elementary systems $\rA$, since a state $\rho_\rA$ can be expressed as $\lambda_\rA I_\rA+T$, with $\lambda_\rA=\frac1{d_\rA}$, and the set of possible traceless $T$ in this case spans the whole $\Trnset_0(\rA)$. Now, let the thesis be true for the types $x,y$. Then, since for every $X\geq0$ on $\sH_x$ there exists $\mu>0$ such that $\mu X\leq \lambda_xI_x$, we have $\con(x)=\pos(\sH_x)$ (and similarly for $y$). Therefore, $\con(x\to y)$ is the cone $\pos(\sH_x\otimes\sH_y)$ of positive operators on $\sH_x\otimes\sH_y$. Moreover, since $\lambda_xI_x\in\Trnset_1(x)$, the deterministic events $R$ in $\Trnset_1(x\to y)$ must satisfy
\begin{align*}
\Tr_x[(I_y\otimes \lambda_xI_x)R]=\lambda_x\Tr_x R=\lambda_yI_y+T,\ T\in\Delta_y
\end{align*}
and thus $\Tr[R]=\lambda_yd_y/\lambda_x$ independently of $R$. This implies that $R=\lambda_{x\to y}I_{x\to y}+T'$ where
\begin{align*}
&\Tr[T']=0,\\
&\lambda_{x\to y}:=\frac{\lambda_y}{\lambda_x d_x}.
\end{align*} 
Finally, the traceless part $T'$ must satisfy
\begin{align*}
\Tr[(S_y\otimes T_x)T']=0,\quad&\forall T_x\in\Delta_x,\ S_y\in\overline\Delta_y,
\end{align*}
where $\overline\Delta_y$, is the complement in the space of traceless operators of $\Delta_y$:
\begin{align*}
\Gamma_y:=\{\lambda I_y|\lambda\in\Reals\},\quad\overline\Delta_y:=(\Gamma_y\oplus\Delta_y)^\perp.
\end{align*}
In other words, 
\begin{align*}
T'\in \Span[&(\Gamma_y\otimes\overline\Delta_x)\oplus(\Delta_y\otimes\Gamma_x)\oplus(\Delta_y\otimes\overline\Delta_x)\\
&\oplus(\overline\Delta_y\otimes\overline\Delta_x)\oplus(\Delta_y\otimes\Delta_x)].
\end{align*}
%\{T\otimes S|T\in \overline \Delta_y^\perp\vee S\in\Delta_x^\perp\}$. Viceversa, if $T'\in \Span\{T\otimes S|T\in \overline\Delta_y^\perp\vee S\in\Delta_x^\perp\}
Thus there exists $\mu>0$ such that $-\mu T'\leq\lambda_{x\to y}I_{x\to y}$, and then $\lambda_{x\to y}I_{x\to y}+\mu T'=:R\geq0$. Clearly, for $X\in\det(x)$ one has
\begin{align*}
\Tr_x[(I\otimes X)R]=\lambda_y I_y+\mu S,
\end{align*}
where $S=\Tr_x[(I\otimes X)T']$. By construction, $\Tr[S]=0$ and $S\in\Delta_y$. Thus, $\Tr_x[(I\otimes X)R]\in\det(y)$, and $R\in \det(x\to y)$. This implies that the operators $T'$ in the decomposition of deterministic events in $\det(x\to y)$ span the whole space 
\begin{align*}
\Delta_{x\to y}=\Span[&(\Gamma_y\otimes\overline\Delta_x)\oplus(\Delta_y\otimes\Gamma_x)\oplus(\Delta_y\otimes\overline\Delta_x)\\
&\oplus(\overline\Delta_y\otimes\overline\Delta_x)\oplus(\Delta_y\otimes\Delta_x)]. 
\end{align*}Consequently, $\lambda_{x\to y}I_{x\to y}\in\Trnset_1(x\to y)$.
%, and this implies that $\con(x\to y)=\pos(\sH_x\otimes\sH_y)$.
\qed

\begin{corollary}
Events of type $x$ generate the full cone $\pos(\sH_x)$ of positive operators on $\sH_x$---that we will denote by $\con(x)$. In formula, $\con(x)=\pos(\sH_x)$
\label{cor:conpos}
\end{corollary}
\Proof Since $\lambda_x I_x\in\det(x)$, and for every $T$ in $\pos(\sH_x)$ there exists $\mu>0$ such that $\mu T\leq\lambda_x I_x$, one has $T\in\con(x)$.\qed
%This is true for types $\rA$, with $\lambda_\rA=1/d_\rA$. Let now the thesis be true for $x,y$. Then, clearly the cone of completely positive maps from the cone $\con(x)$ to the cone $\con(y)$ is in bijective correspondence through the Choi isomorphism with a cone of positive operators on $\sH_x\otimes\sH_y$. Let us now expand $.$
%\qed
\begin{corollary}
It is $X\in\det(x)$ iff $X^T\in\det(X)$.
\end{corollary}

From now on, given a deterministic event $X\in\det(x)$, we will denote the traceless operator in the decomposition \eqref{eq:decompdet} of $X$ as $T_X$. Clearly, $T_{X^T}=T_X^T$.

One can now
easily prove the following lemma
\begin{lemma}\label{lem:types}
  An element $X\in\con(x)$ is a deterministic event of type $x$ if and only if
  \begin{equation}\label{eq:normconst}
    \Tr[XY]=1,
  \end{equation}
  for every $Y\in\Trnset_1(\overline x)$.
\end{lemma}
\Proof Necessity. The only element of $\Trnset_1(\rI)$ is 1.  Then, since $\overline x=x\to \rI$, saying that $Y\in\Trnset_1(\overline x)$ is equivalent to saying that $Y\in\con(x)$ and
$Y$ satisfies Eq.~\eqref{eq:normconst} for every $X\in\Trnset_1(x)$. Then, for every $X\in\det(x)$, one must have that for every $Y\in\det(\overline x)$ Eq.~\eqref{eq:normconst} holds. In other words, satisfying Eq.~\eqref{eq:normconst} for every $Y\in\det(\overline x)$ is a necessary condition for $X$ to be
deterministic.

Sufficiency. Let $X\in\con(x)$, and suppose that  Eq.~\eqref{eq:normconst} is satisfied for every $Y\in\Trnset_1(\overline x)$. By lemma \ref{lem:firstrecpro} one has $\lambda_{\overline x}I_x\in\det(\overline X)$, and then
\begin{align}
\Tr[X]\lambda_{\overline x}=1.
\label{eq:normal}
\end{align}
Since by lemma \ref{lem:firstrecpro} one also has that $Y\in\det(\overline x)$ has the form $Y=\lambda_{\overline x}I_x+T_Y$ with $T_Y\in\Delta_{\overline x}$, by Eq.~\eqref{eq:normal} one has $\Tr[XT_Y]=0$, for every $Y$, namely 
\begin{align*}
\Tr[XT]=0,\ \forall T\in\overline{\Delta}_{x}.
\end{align*}
This implies that
\begin{align*}
X=\lambda_x I_x+T_X,
\end{align*}
where $\Tr[T_X]=\Tr[T_X T]=0$ for all $T\in\overline\Delta_{x}$, namely $T_X\in\Delta_x$. By lemma \ref{lem:firstrecpro} we have then $X\in\Trnset_1(x)$.
%
%If $X\in\Trnset(x)$, one has $ X+Z=R$ for some $Z\in\Trnset(x)$ and $R\in\Trnset_1(x)$. Then it must be $Y*Z=0$ for every $Y\in\Trnset_1(\overline x)$.
%Now, since for every $W\in\Trnset(\overline x)$ there exist
%$W'\in\Trnset(\overline x)$ and $Y\in\Trnset_1(\overline x)$ such that $W+W'=Y$,
%this implies that $W*Z=0$ for every $W\in\Trnset(\overline x)$, and finally
%this implies \footnote{Indeed, the cone $\con(\overline x)$ is by construction a
%  separating cone of linear functionals on $\Trnset_{\mathbb R}(x)$.}. 
%  $Z=0$, namely $X\in\Trnset_1(x)$ is
%deterministic.
\qed

\begin{corollary}\label{cor:barbar}
  The following identity holds 
  \begin{equation}
    \overline{\overline x}=x.
  \end{equation}
\end{corollary}

Finally, the following lemma holds for the space $\Delta_x$

\begin{lemma}
One has $X\in\Delta_x$ if and only if $X=\lambda(X_1-X_2)$ with $X_1,X_2\in\det(x)$. 
\label{lem:diff}
\end{lemma}
\Proof Let $X\in\Delta_x$, and define $X_1:=\lambda_xI_x+\mu X$ with $\mu>0$ such that $X_1\geq0$. Then consider $X_2:=\lambda_xI_x$. Thus, by lemma \ref{lem:firstrecpro} one has $X_1,X_2\in\det(x)$, and clearly $X=1/\mu(X_1-X_2)$. Viceversa, let $X=X_1-X_2$ for $X_1,X_2\in\det(x)$. By lemma \ref{lem:firstrecpro} one has $X_1-X_2=T_{X_1}-T_{X_2}\in\Delta_x$.\qed

As a consequence of the above results, it is easy to prove the following lemma.

\begin{lemma}
One has $\Delta_{\overline x}=\overline\Delta_x$.
\end{lemma}

\Proof  By lemma \ref{lem:diff}, for $Y\in\Delta_{\overline x}$ one has $Y=\lambda(Y_1-Y_2)$ with $Y_1,Y_2\in\det(\overline x)$. This implies that $\Tr[YX]=0$ for all $X\in\det(x)$. Since $\lambda_xI_x\in\det(x)$, one has $\Tr[Y]=0$ and  $\Tr[YT]=0$ for all $T\in\Delta_x$. Thus, $\Delta_{\overline x}\subseteq\overline \Delta_x$. Finally, by corollary \ref{cor:barbar} one also has $\Delta_{\overline x}=\Delta_{\overline{\overline{\overline x}}}\subseteq\overline\Delta_{\overline{\overline x}}=\overline\Delta_x$. 
\qed
The spaces $\Delta_x$ provide a simple and useful decomposition of the real space $\herm(\sH_x)$ of selfadjoint operators on $\sH_x$. Indeed, one has
\begin{align*}
\herm(\sH_x)&=\Gamma_x\oplus\Delta_x\oplus\Delta_{\overline x}\\
&=\Gamma_x\oplus\Delta_x\oplus\overline\Delta_{x}.
\end{align*}
From the proof of lemma \ref{lem:firstrecpro}, one has
\begin{align}
\Delta_{x\to y}=&(\Gamma_x\otimes\overline\Delta_y)\oplus(\overline\Delta_x\otimes\overline\Delta_y)\nonumber\\
&\oplus (\Gamma_x\otimes{\Delta_y})\oplus(\overline\Delta_x\otimes\Gamma_y)\oplus(\overline\Delta_x\otimes{\Delta_y})\nonumber\\
&\oplus(\Delta_x\otimes{\Gamma_y})\oplus(\Delta_x\otimes\Delta_y).
\label{eq:deltaxtoy}
\end{align}
Moreover, by direct evaluation one has
\begin{align}
\Delta_{x\otimes y}=(\Delta_x\otimes\Gamma_y)\oplus(\Gamma_x\otimes\Delta_y)\oplus(\Delta_x\otimes\Delta_y).
\label{eq:deltaxoty}
\end{align}

We will now prove that $\otimes$ is
associative. For this purpose we need the following lemma.
%\begin{lemma}\label{lem:tenstens}
%  The cone $\con(x\otimes y)$ of events is the cone of positive operators on $\sH_x\otimes\sH_y$.
%\end{lemma}
%\Proof Let $F\in\Trnset(x\otimes y)$. This implies that $F*G\in\con(I)$ for every $G\in$

\begin{lemma}\label{lem:afftens}
  The set of deterministic events $\Trnset_1(x\otimes y)$ is the
  intersection $\con(x\otimes y)\cap\mathsf A$ of the cone
  $\con(x\otimes y)$ with
  \begin{equation}\label{eq:afftens}
    \mathsf A:=\aff\{X\otimes Y|X\in\Trnset_1(x), Y\in\Trnset_1(y)\},
  \end{equation}
  $\aff(S)$ denoting the affine span of $S$.
\end{lemma}
\Proof By definition, $F\in \Trnset_1(x\to \overline y)$ if and only if $F\geq 0$ and 
$\Tr[Y\Tr_x[F(I\otimes X^T)]=\Tr[F(Y\otimes X^T)]=1$ for every $X\in\Trnset_1( x)$ and $Y\in
\Trnset_1(y)=\Trnset_1(\overline{\overline y})$. Thus, by lemma \ref{lem:types}, every operator $X\otimes Y$ with $X\in\Trnset_1(x)$ and $Y\in\Trnset_1( y)$ is an element of
$\det(\overline{x\to\overline y})=\Trnset_1(x\otimes y)$, and clearly the same holds for every $G\in(\mathsf A\cap\con(x\otimes y))$.
Thus, $(\mathsf A\cap\con(x\otimes y))\subseteq \Trnset_1(x\otimes y)$,
%On the other hand, 
%for $F\in\Trnset(x\to\overline y)$ one has $F\in\det(x\to\overline y)$ if and only if $F*G=1$ for every $G\in(\mathsf A\cap\con(x\otimes y))$, and also by lemma \ref{lem:types} $F\in\det(x\to\overline y)$ if and only if $F*X=1$ for every $X\in \det(\overline{x\to\overline y})$. Thus, for $F\in\Trnset(x\to\overline y)$ one has
%\begin{align*}
%&F*G=1&&&\forall G\in\mathsf A\cap\con(x\otimes y)\\
%&&\Leftrightarrow&\\ 
%&F*X=1&&&\forall X\in\det(\overline{x\to\overline y}).
%\end{align*}
%Suppose now  
and defining $\mathsf A_+:=\Span(\{Z-\lambda_{\overline{x\to\overline y}}I_{\overline{x\to\overline y}}|Z\in\mathsf A\cap\con(x\otimes y)\})$, by lemma \ref{lem:firstrecpro} we have that 
\begin{align*}
\Delta_{\overline {x\to\overline y}}\supseteq \set A_+.
\end{align*}
On the other hand, suppose that the above inclusion is strict.
Then one has
\begin{align*}
\overline\Delta_{\overline {x\to\overline y}}\subset [\Gamma_{x\to y}\oplus\set A_+]^\perp,
\end{align*}
which implies the existence of a traceless $T\in\Delta_{\overline {x\to\overline y}}\cap\mathsf A_+^\perp$. If we now form the positive operators
\begin{align*}
&Z:=\lambda_{x\to\overline y}I_{x\to\overline y}+\mu T,\\
&W:=\lambda_{\overline{x\to\overline y}}I_{\overline{x\to\overline y}}+\nu T
\end{align*}
choosing suitable non-null reals $\mu,\nu$, on one hand we have
\begin{align*}
\Tr[Z(X\otimes Y)]=1,\quad \forall X\in\det(x),\ Y\in\det(y),
\end{align*}
which implies $Z\in\det(x\to\overline y)$, but also
\begin{align*}
W\in\det(\overline{x\to\overline y}),
\end{align*}
by lemma \ref{lem:firstrecpro}. Finally, this leads to the following identity
\begin{align*}
\Tr[ZW]=1+\mu\nu\Tr[T^2]\neq1,
\end{align*}
in contradiction with lemma \ref{lem:types}. Then it must be 
\begin{align*}
\Delta_{\overline {x\to\overline y}}= \set A_+,
\end{align*}
and
\begin{align*}
(\mathsf A\cap\con(x\otimes y))= \Trnset_1(x\otimes y).
\end{align*}
\qed

\begin{corollary}\label{cor:comm}
  $x\otimes y= y\otimes x$
\end{corollary}

As a consequence of Corollary \ref{cor:comm} $\overline{x\to\overline
  y}=\overline{y\to\overline x}$.  Substituting $y$ by $\overline y$ we obtain
the following identity
\begin{equation}
  x\to y= \overline y\to\overline x.
\end{equation}

It is now possible to prove that every event type $x\to y$ is
equivalent to a type $x'\to(\rA\to\rB)$. The general notion behind
this result is known in computer science as {\em Currying}---more
precisely its opposite, {\em uncurrying}---which we clarify in the
next lemmas.

% \begin{lemma} Functions $f$ of type $f:(x,y)\to z$, with $x\in\rX$,
%   $y\in\rY$ anf $z\in\rZ$ are in one-to-one correspondence with
%   functions $F$ of type $F:x\to(y\to z)$.
% \end{lemma}

% \Proof The correspondence $\cu$ is defined as follows
% \begin{align}
%   &[\cu(f)(x)](y):=f(x,y),\nonumber\\
%   &[\cu^{-1}(F)](x,y):=[F(x)](y).
% \end{align}
% It is trivially verified that $\cu^{-1}([\cu(f)])=f$ and
% $\cu([\cu^{-1}(F)])=F$. Indeed
% \begin{align}
%   &\cu^{-1}([\cu(f)])(x,y)=[\cu(f)(x)](y)=f(x,y),\nonumber\\
%   &\{\cu([\cu^{-1}(F)])(x)\}(y)=[\cu^{-1}(F)](x,y)=[F(x)](y).
% \end{align}
% This completes the proof.\qed

%A similar lemma can be proved for types, as follows.
\begin{lemma}\label{lem:uncurryass}
  Associativity of $\otimes$, namely the identity
  \begin{equation}\label{eq:ass}
    (x\otimes y)\otimes z=x\otimes(y\otimes z) \quad\forall x,y,z,
  \end{equation}
  is equivalent to the {\em uncurrying identity}
  \begin{equation}\label{eq:uncu}
    x\to(y\to z)=(x\otimes y)\to z \quad\forall x,y,z,
  \end{equation}
\end{lemma}
\Proof Let us suppose that Eq.~\eqref{eq:ass} holds. By definition, we
have then $\overline{(x\otimes y)\to\overline z}=\overline{x\to(y\to\overline
  z)}$, $\forall x,y,z,$, namely (substituting $z$ for $\overline z$)
\begin{equation}
    (x\otimes y)\to z=x\to(y\to z) \quad\forall x,y,z.
\end{equation}
Conversely, if $(x\otimes y)\to z=x\to(y\to z)$ $\forall x,y,z$, then
$\overline{(x\otimes y)\otimes\overline z}=\overline{x\otimes(y\otimes\overline
  z)}$, namely
\begin{equation}
  (x\otimes y)\otimes z= x\otimes(y\otimes z).
\end{equation}
\qed

We now prove associativity of $\otimes$, which then trivially implies
the uncurrying identity.

\begin{lemma}\label{lem:ass}
  For every triple $x,y,z$, $(x\otimes y)\otimes z=x\otimes(y\otimes
  z)$.
\end{lemma}
\Proof Since $\con([(x\otimes y)\otimes z])=\con([x\otimes(y\otimes
z)])$, it is sufficient to prove that $\det([(x\otimes y)\otimes
z])=\det([x\otimes(y\otimes z)])$. For this purpose, we define
\begin{align}
  &\mathsf A:=\aff\{W\otimes Z|W\in\det(x\otimes y),\ Z\in\det(z)\},\nonumber\\
  &\mathsf B:=\aff\{X\otimes T|X\in\det(x),\ T\in\det(y\otimes z)\}.
\end{align}
and we remind Eq.~\eqref{eq:afftens}, which implies
\begin{align}
  &\det([(x\otimes y)\otimes z])=\mathsf A\cap\pos(\sH_x\otimes \sH_y\otimes \sH_z),\nonumber\\
  &\det([x\otimes(y\otimes z)])=\mathsf B\cap\pos(\sH_x\otimes \sH_y\otimes \sH_z).
\end{align}
We will now prove that $\mathsf A=\mathsf C$, where $C:=\aff\{X\otimes
Y\otimes Z|X\in\det(x),\ Y\in\det(y),\ Z\in\det(z)\}$. It is trivial
to verify that $\mathsf A\subseteq\mathsf C$. Consider now a general
element $V\in\mathsf C$. By definition there exist real coefficients
$\{a_{p,q,r}\}_{(p,q,r)\in\samp P\times\samp Q\times\samp
  R}\subseteq\mathbb R$ and elements $\{X_p\}_{p\in\samp P}\subseteq
\det (x)$, $\{Y_q\}_{q\in\samp Q}\subseteq \det (y)$ and
$\{Z_r\}_{r\in\samp R}\subseteq\det(z)$, such that
\begin{align}
  &\sum_{(p,q,r)\in\samp P\times\samp Q\times\samp
    R}a_{pqr}=1,\nonumber\\
  &\sum_{(p,q,r)\in\samp P\times\samp Q\times\samp
    R}a_{pqr}X_p\otimes Y_q\otimes Z_r=V.
\end{align}
Let $c_r:=\sum_{p'q'}a_{p'q'r}$, and $b^r_{pq}:=a_{pqr}/c_r$. It is clear that $\sum_rc_r=1$, and $\sum_{pq}b^r_{pq}=1$ for every $r$. Thus we have
\begin{align}
V&=\sum_{pqr}c_rb^r_{pq}X_p\otimes Y_q\otimes Z_r\nonumber\\
&=\sum_rc_rT_r\otimes Z_r,
\end{align}
where $T_r:=\sum_{pq}b^r_{pq}X_p\otimes Y_q\in\det(x\otimes y)$. This proves that $V\in\mathsf A$, and then $\mathsf C\subseteq\mathsf A$.
A similar proof clearly holds also for $\mathsf B$ thus providing the
thesis, since $\mathsf A=\mathsf C=\mathsf B$.\qed

\begin{corollary}\label{cor:uncurtyp}
  For every triple $x,y,z$ the following type equality holds $x\to (yx\to
  z)=(x\otimes y)\to z$.
\end{corollary}

Every new type $x$ in the hierarchy comes from a couple $y,z$ through
one of the two type compositions. This allows us to introduce a binary
relation on types as follows
\begin{align}
    &y\preceq x\ \mbox{if and only if there exists}\ z,\\
    &x=y\otimes z \ \mbox{or}\ x=y\to z\ \mbox{or}\ y=z\to x
\end{align}
Using the definition of $\otimes$, we can restate our definition as follows
\begin{definition}\label{def:prec}
  We say that $y$ is a parent of $x$, and denote it as $y\preceq x$,
  if there exists $z$ such that one of the following conditions
  holds
  \begin{enumerate}
  \item $x=y\otimes z$,
  \item $x=\overline y\otimes z$,
  \item $x=\overline{y\otimes \overline z}$,
  \item $x=\overline{\overline y\otimes z}$.
  \end{enumerate}
\end{definition}

\begin{definition}
Let us define the binary relation $R$ between types such that $x R y$ if $x\preceq y$ and $y\preceq x$.
\end{definition}

\begin{lemma}\label{lem:coool}
One has $xRy$ iff $x=y$ or $x=\overline y$.
\end{lemma}

\Proof Since $\dim([\Trnset_{\mathbb R}(x)])=\dim([\Trnset_{\mathbb R}(\overline x)])$ and $\dim([\Trnset_{\mathbb R}(x\otimes z)])=\dim([\Trnset_{\mathbb R}(x)])\dim([\Trnset_{\mathbb R}(z)])$, if $x\prec y$ one necessarily has $\dim([\Trnset_{\mathbb R}(x)])\leq \dim([\Trnset_{\mathbb R}(y)])$. Thus, if $x\preceq y$ and $y\preceq x$ one has $\dim([\Trnset_{\mathbb R}(x)])=\dim([\Trnset_{\mathbb R}(y)])$. Thus, in all the four cases of definition \ref{def:prec}, it must be $\dim([\Trnset_{\mathbb R}(z)])=1$, namely $z=I$. This finally implies either $x=y$ or $x=\overline y$. \qed

\begin{corollary}
The relation $R$ is reflexive, symmetric and transitive.
\end{corollary}

If we quotient the set of types by the relation $R$, the equivalence classes inherit a relation $\preceq$ defined as
\begin{align}
[x]\preceq [y]\ \Leftrightarrow\ x\preceq y.
\end{align}
One can easily verify that the relation $\preceq$ between equivalence classes is well defined. Indeed, by lemma \ref{lem:coool} one has $[x]=\{x,\overline x\}$. Thus, if $x\preceq y$ one also has $\overline x\preceq y$, $x\preceq \overline y$ and $\overline x\preceq\overline y$.

\begin{lemma} 
The relation $\preceq$ between equivalence classes is reflexive and antisymmetric.
\end{lemma}
\Proof Reflexivity is simply proved, because for any $x$ we have $x\otimes \rI=x$. 
%Suppose that $x\preceq y$ and $y\preceq x$. Then there are nine possibilities:
%\begin{enumerate}
%\item
%$x=y\otimes z$ and $y=x\otimes z'$. Then $x=x\otimes (z\otimes z')$.
%\item
%$x=y\otimes z$ and $\overline y=x\otimes z'$. Then $y=\overline y\otimes(z\otimes z')$.
%\item
%$x=y\otimes z$ and $\overline y=\overline x\otimes z'$. Then $.$
%\end{enumerate}
Suppose now that $[x]\preceq [y]$ and $[y]\preceq [x]$. Then $x\preceq y$ and $y\preceq x$, namely $xRy$ and thus
$[x]=[y]$. \qed

In the following we will denote by $\preceq$ the {\em transitive closure} of the relation $\preceq$ between equivalence classes. It is then clear that $\preceq$ is a partial ordering in the quotient of types modulo $R$.

Since every type $x$ is obtained from elementary types by subsequent applications of $\otimes$ and $\overline{\phantom{o}}$, we can prove the property $P$ of types by induction with respect to the ordering $\preceq$ by proving it for every elementary type $A$, namely $P(A)=1$, and then proving that $P(x)=1\Rightarrow P(\overline x)=1$, and $P(x)=P(y)=1\Rightarrow P(x\otimes y)=1$. 

The above induction technique will be used to prove the main result of the paper in section \ref{s:charact}. 

Finally, we now define the notion of {\em intersection} of types.
\begin{definition}
Let $z$ be a type such that $\sH_z=\sH_x=\sH_y$ for two types $x,y$, and $\det(z)=\det(x)\cap\det(y)$. We say that the type $z$ is the {\em intersection} of types $x$ and $y$, and write $z=x\cap y$.
\label{def:capdets}
\end{definition}
This definition bears the following elementary consequences.
\begin{lemma}
The type $z$ is the intersection of $x,y$ if and only if $\sH_z=\sH_x=\sH_y$, $\lambda_z=\lambda_x=\lambda_y=:\lambda$, and $\Delta_z=\Delta_x\cap\Delta_y$.
\end{lemma}
\Proof By definition, if $z=x\cap y$ it must be $\lambda_z=\lambda_x=\lambda_y$. Moreover, for every $Z\in\det(z)$ one has
\begin{align*}
Z=\lambda I_z+T_Z,
\end{align*}
with $T_Z\in\det(x)\cap\det(y)$. Thus, $\Delta_z\subseteq\Delta_x\cap\Delta_y$. Moreover, if $T_Z\in\Delta_x\cap\Delta_y$ then clearly $Z:=\lambda I+T_Z\in\det(x)\cap\det(y)$, namely $Z\in\det(z)$. Then $\Delta_x\cap\Delta_y\subseteq\Delta_z$, and finally this implies $\Delta_x\cap\Delta_y=\Delta_z$. Conversely, let $\sH_x=\sH_y=\sH_z$, $\lambda_x=\lambda_y=\lambda_z$, and $\Delta_z=\Delta_x\cap\Delta_y$. Then if $Z\in\det(z)$ it clearly belongs to both $\det(x)$ and $\det(y)$ by virtue of lemma \ref{lem:firstrecpro}.\qed
\begin{lemma}
Let $z=x\cap y$. One has $\Delta_{\overline z}=\Span(\Delta_{\overline x}\cup\Delta_{\overline y})$.
\label{lem:complemint}
\end{lemma}
\Proof Let $T\in\overline\Delta_{x}=\Delta_{\overline x}$. Then clearly $\Tr[TT_W]=0$ for every $T_W\in\Delta_{z}$, and the same argument holds for $T\in\overline\Delta_{y}=\Delta_{\overline y}$. Thus, we have that $\Delta_{\overline x}\subseteq\Delta_{\overline z}$ and $\Delta_{\overline y}\subseteq\Delta_{\overline z}$. Thus, $\Span(\Delta_{\overline x}\cup\Delta_{\overline y})\subseteq\Delta_{\overline z}$. Suppose now that there is $0\neq T\in\Delta_{\overline z}$, and $T\not\in(\Delta_{\overline x}\cup\Delta_{\overline y})$. Then the component of $T'\neq0$ of $T$ in $\Span(\Delta_{\overline x}\cup\Delta_{\overline y})^\perp\cap \Delta_{\overline z}$ satisfies
\begin{align*}
\Tr[T'T_W]&=0,\quad\forall T_W\in\Delta_z,\\
\Tr[T'T_{\overline X}]=\Tr[T'T_{\overline Y}]&=0,\quad\forall T_{\overline X}\in\Delta_{\overline x},\ T_{\overline Y}\in\Delta_{\overline y}.
\end{align*}
The equalities in the second line imply that $T'\in\Delta_x\cap\Delta_y$, while the one on the first implies $T'\in\Delta_{\overline z}$. Thus, we have $T'\in\Delta_z\cap\Delta_{\overline z}=\{0\}$, contrarily to the hypothesis. Then, it must be $$\Span(\Delta_{\overline x}\cup\Delta_{\overline y})=\Delta_{\overline z}$$
%First of all, by the proof of lemma \ref{lem:firstrecpro}, $\lambda_x=\lambda_y$ implies that $\lambda_{\overline x}=\lambda_{\overline y}=:\lambda$. Moreover, by the same lemma one has $\con(\overline z)=\con(\overline x)=\con(\overline y)$. This means that 
%\begin{align*}
%W\in\det(\overline{z})\ \Leftrightarrow\ W=\lambda I_{\overline{z}}+T_W,
%\end{align*}
%where $\Tr[T_WT_X]=\Tr[T_WT_Y]\in\Delta_{\overline{z}}$, and by lemma \ref{lem:complemint} it is $T_W=T_{\overline x}+T_{\overline y}$ with $T_z\in\Delta_z$ for $z=\overline x,\overline y$. This implies that there exists $\mu$ such that 
\qed

Moreover, we have the two following important lemmas.
\begin{lemma}
For every pair of types $x,y$ one has 
\begin{align}
\det(\overline{x\cap y})=\con(\overline{x\cap y})\cap\aff[\det(\overline x)\cup\det(\overline y)].
\end{align}
\label{lem:compcapdet}
\end{lemma}

\Proof Let $T\in\det(\overline{x\cap y})$. Then
\begin{align*}
T=\lambda_{\overline{x\cap y}}+Z,\quad Z\in\Span(\Delta_{\overline x}\cup\Delta_{\overline y}),
\end{align*}
namely $Z=\alpha \overline X+\beta \overline Y$ with $\overline X\in\Delta_{\overline x}$ and $\overline Y\in\Delta_{\overline y}$. By choosing a suitably large $\theta$, it is always possible to have
\begin{align*}
A:=&\lambda_{\overline{x\cap y}}+\frac\alpha\theta\overline X\geq0,\\
B:=&\lambda_{\overline{x\cap y}}+\frac\beta{(1-\theta)}\overline Y\geq0.
\end{align*}
Thus, $T=\theta A+(1-\theta)B$. Now, clearly $A\in\det(\overline x)$ and $A\in\det(\overline y)$.\qed

\begin{lemma}
For every pair of types $x,y$ and every $z$, one has
\begin{align}
&(x\cap y)\otimes z=(x\otimes z)\cap(y\otimes z).
\label{eq:captens}
\end{align}
\label{lem:captens}
\end{lemma}
\Proof First of all, we observe that by Corollary \ref{cor:conpos}, we have $\con(x\cap y)=\con(x)=\con(y)$. Then, $\con[(x\cap y)\otimes z]=\con(x\otimes z)=\con(y\otimes z)$. Moreover, by definition \ref{def:capdets} along with lemma \ref{lem:afftens}, we have
\begin{align*}
\det[&(x\cap y)\otimes z]=\con[(x\cap y)\otimes z]\\
&\cap\aff\{W\otimes Z|W\in\det(x\cap y),Z\in\det(z)\}\\
=&\con[(x\cap y)\otimes z]\\
&\cap\aff\{W\otimes Z|W\in\det(x),W\in\det(y),Z\in\det(z)\}\\
=&\con(x\otimes z)\cap\aff\{W\otimes Z|W\in\det(x),Z\in\det(z)\}\\
&\cap\con(y\otimes z)\cap\aff\{W\otimes Z|W\in\det(y),Z\in\det(z)\}\\
=&\det(x\otimes z)\cap\det(y\otimes z).
\end{align*}
This implies the thesis.\qed
\begin{lemma}
For every pair of types $x,y$ and every $z$, one has
\begin{align}
&\Delta_{(x\cap y)\to z}=\Span(\Delta_{x\to z}\cup\Delta_{y\to z})
\end{align}
\label{lem:deltacapto}
\end{lemma}
\Proof Let us remind that $(x\cap y)\to z=\overline{(x\cap y)\otimes \overline z}$, and thus
\begin{align*}
\Delta_{(x\cap y)\to z}=\overline \Delta_{(x\cap y)\otimes \overline z}.
\end{align*}
Now, by Lemma \ref{lem:captens} we have
\begin{align*}
\Delta_{(x\cap y)\otimes \overline z}=\Delta_{(x\otimes\overline z)\cap (y\otimes \overline z)},
\end{align*}
and finally by Lemma \ref{lem:complemint} we have
\begin{align*}
\Delta_{(x\cap y)\to z}=&\Span(\Delta_{\overline{x\otimes\overline z}}\cup\Delta_{\overline{y\otimes\overline z}})\\
=&\Span(\Delta_{x\to z}\cup\Delta_{y\to z}).
\end{align*}
\qed

\begin{corollary}
For every pair of types $x,y$ and every $z$, one has\begin{align*}
\det[(x\cap y)\to z]=&\con[(x\cap y)\to z]\\
&\cap\aff[\det(x\to z)\cup\det(y\to z)].
\end{align*}
\label{cor:detaffcap}
\end{corollary}

\Proof The result is an immediate consequence of Lemma \ref{lem:deltacapto} along with Lemma \ref{lem:compcapdet}.\qed

\section{Characterisation of general maps}\label{s:charact}

In the following we will prove results that depend on the structure of a type $x$ rather than on the dimension of the specific elementary systems $\rA_i$ that compose it. For example, we will treat on the same footing transformations $\rA_0\to\rB_0$ and $\rA_1\to\rB_1$, even if $d_{\rA_0}\neq d_{\rA_1}$ or $d_{\rB_0}\neq d_{\rB_1}$. For this purpose of the present section, it is convenient to introduce a notation which is at the same time insightful and efficient. Given a Hilbert space $\sH_x=\sH_n\otimes\sH_{n-1}\otimes\ldots\otimes\sH_{0}$, one can expand any operator on $\sH_x$ on the basis
$\{S_{\bi}=S^{(n)}_{i_n}\otimes S^{(n-1)}_{i_{n-1}}\otimes\ldots\otimes S^{(0)}_{i_0}\}$, where $S^{(j)}_0:=I_{\sH_j}$, and for every $j$ it is $\Tr[S^{(j)}_{l}]=0$ for $l>0$. In the following we will denote $T^{(j)}_l:=S^{(j)}_l$ for $l>0$. An important role in our analysis is played by those special subspaces of $\herm(\sH_x)$ having the following property: they are spanned by a subset of $\{S_{\bi}\}$ such that for every $j$, either all the $S_\bi$ in the subset have $S^{(j)}_{i_j}=I_{\sH_j}$, or they all have $\Tr[S^{(j)}_{i_j}]=0$. As an example, let $\sH_x=\sH_1\otimes \sH_0$. Then we have four subspaces of interest:
\begin{align*}
\sL_{00}:=&\Span(\{T_i\otimes T_j\})\\
\sL_{01}:=&\Span(\{T_i\otimes I\})\\
\sL_{10}:=&\Span(\{I\otimes T_j\})\\
\sL_{11}:=&\Span(\{I\otimes I\}).
\end{align*}
In the general case, we will define the space $\sL_{\vb}$, where $\vb$ is a string of bits of length $n+1$, as follows: $\sL_{\vb}$ is the largest subspace spanned by $S_\bi$'s such that for all those values of $j$ for which $b_j=1$ one has $i_j=0$, i.e.~$S^{(j)}_{i_j}=I_{\sH_i}$, while for all those values of $j$ for which $b_j=0$ one has $i_j>0$, i.e.~$\Tr[T^{(j)}_{i_j}]=0$. 

As a consequence of the definition, one has the following remarkable identity for a string $\vb=\vb_1\vb_0$
\begin{align}
\sL_{\vb_1\vb_0}=\sL_{\vb_1}\otimes\sL_{\vb_0}.
\label{eq:juxtel}
\end{align}

Notice that the notation $\sL_\vb$ is not reminiscent of the particular dimensions of spaces $\sH_j$. This is due to the fact that the dimensions play almost no role in the structure theorems that we prove in the following.

What is crucial about the mentioned subspaces is that it is particularly easy to figure out their intersection and their sum. Indeed, let $\Pi_\vb$ denote the projection on the subspace $\sL_\vb$ of $\herm(\sH_x)$. Then one can prove the following lemma.

\begin{lemma}
Let $\Pi_{\vb_0}$ and $\Pi_{\vb_1}$ be the projections on the subspaces $\sL_{\vb_0}$ and $\sL_{\vb_1}$, respectively. Then
\begin{align}
[\Pi_{\vb_0},\Pi_{\vb_1}]=0.
\end{align}
\end{lemma}

\Proof The statement is trivial when $\vb_0=\vb_1$. Let us then focus on the case $\vb_0\neq\vb_1$. One easily realises that every element of the basis of $\sL_{\vb_0}$ is orthogonal to every element of the basis of $\sL_{\vb_1}$ in the Hilbert-Schmidt sense. Indeed, $\vb_0\neq\vb_1$ implies that there exists some $j$ such that $(b_0)_j\neq(b_1)_j$. Let us suppose without loss of generality that $(b_0)_j=0$ and $(b_1)_j=1$. Then we have
\begin{align*}
\Tr[&(S^{(n)}_{i_n}\otimes S^{(n-1)}_{i_{n-1}}\otimes\ldots I_j\otimes\ldots \otimes S^{(0)}_{i_0})\\
&\times(S'^{(n)}_{l_n}\otimes S'^{(n-1)}_{l_{n-1}}\otimes\ldots T^{(j)}_{l_j}\otimes\ldots \otimes S'^{(0)}_{l_0})]\\
=&\Tr[S^{(n)}_{i_n}S'^{(n)}_{l_n}]\ldots\Tr[T^{(j)}_{l_j}]\ldots\Tr[S^{(0)}_{i_0})S'^{(0)}_{l_0})]\\
=&k\Tr[T^{(j)}_{l_j}]=0.
\end{align*}
This implies that two subspaces $\sL_{\vb_0}$ and $\sL_{\vb_1}$ are orthogonal, and then the thesis follows.\qed

\begin{corollary}
The sum $\sL_{\vb_1}+\sL_{\vb_2}$ for $\vb_1\neq\vb_2$ is a direct sum $\sL_{\vb_1}\oplus \sL_{\vb_2}$.
\label{cor:selop}
\end{corollary}

\begin{lemma}
Let $J\subseteq\{0,1\}^N$, and $\sL$ be the following direct sum of spaces $\sL_{\vb}$
\begin{align}
\sL=\bigoplus_{\vb\in J}\sL_{\vb} 
\label{eq:sel}
\end{align}
Then its orthogonal complement $\sL^\perp$ is the space
\begin{align}
\sL^\perp=\bigoplus_{\vb\in \overline J}\sL_{\vb},
\label{eq:selp}
\end{align}
where $\overline J:=\{0,1\}^N\setminus J$.
\label{lem:oplus}
\end{lemma}

\Proof Since $\herm(\sH_{x})=\sL_{00\ldots 0}\oplus\sL_{00\ldots 1}\oplus\ldots\oplus\sL_{11\ldots1}$, the set $J\subseteq\{0,1\}^n$ of binary strings identifies the direct sum
\begin{align*}
\herm(\sH_x)=\sL\oplus\sL^\perp,
\end{align*}
as in Eqs.~\eqref{eq:sel} and \eqref{eq:selp}.
\qed

The first observation that we make is that for every type $x$, the space $\Gamma_x=\{\lambda_x I_x|\lambda\in\Reals\}$ coincides with the space $\sL_{\bvec 1}$, with $\bvec 1=11\ldots 1$, i.e.
\begin{align}
\Gamma_x=\sL_\bvec 1.
\label{eq:gammasl}
\end{align}

We then prove the following lemma.
\begin{lemma}
Let $x$ and $y$ be two type classes. Types in the class $x\otimes y$ can be characterised by the following identity
\begin{align}
\Delta_{x\otimes y}=(\sL_\bvec 1\otimes\Delta_y)\oplus(\Delta_x\otimes\Delta_y)\oplus(\Delta_x\otimes\sL_\bvec 1).
\label{eq:deltatens}
\end{align}
\label{lem:deltatens}
\end{lemma}
\Proof Eq.~\eqref{eq:deltatens} is just a consequence of Eq.~\eqref{eq:deltaxoty}.\qed

The following theorem shows that the space $\Delta_x$ corresponding to a type $x$ is indeed a direct sum of spaces $\sL_\vb$. This result is crucial for the remainder of the section.

\begin{theorem}
The space $\Delta_x$ is a direct sum of spaces $\sL_\vb$.
\end{theorem}

\Proof The thesis holds for elementary systems $x=\rA$, since the normalisation of a state is
\begin{align}
\Tr[\rho]=1,
\end{align}
which implies $\rho\in\sL_{1}\oplus\sL_{0}$, thus $\Delta_\rA=\sL_0$. We now prove the general statement by induction. Suppose that the statement is true for types $x,y$. Then, by Eq.~\eqref{eq:gammasl} and lemma \ref{lem:oplus} also $\overline\Delta_{y}$ is a direct sum of spaces $\sL_\vb$. Finally, by lemma \ref{lem:deltatens} and Eq.~\eqref{eq:juxtel}, we have that also $\Delta_{x\otimes y}$ is a direct sum of spaces $\sL_\vb$.\qed 

\subsection{Review on combs}

A particularly relevant sub-hierarchy, that was studied extensively in Refs.~\cite{comblong,tesibis}, is that of {\em combs}, given by the following recursive definition
\begin{definition}
\begin{enumerate}
\item
The type $1_{01}$ of 1-combs on $\sH_{\rA_1}\otimes\sH_{\rA_0}$ is $\rA_0\to \rA_1$. The set $\det(1_{01})$ of deterministic 1-combs on $\sH_{\rA_1}\otimes\sH_{\rA_0}$ is the set of Choi operators of channels in $\rA_0\to\rA_1$.
\item
The type $\rn_{01\ldots (2n-1)}$ of $\rn$-combs on $\sH_{\rA_{2n-1}}\otimes\sH_{\rA_{2n-2}}\otimes\ldots\otimes\sH_{\rA_0}$ is 
\begin{align}
(\rn-1)_{1\ldots (2n-2)}\to 1_{0(2n-1)}. 
\end{align}
The elements $R$ of the set $\det(\rn_{01\ldots (2n-1)})$ are Choi operators of CP maps that transform elements of $\det(\rn_{1\ldots (2n-2)})$ to elements of $\det(1_{0(2n-1)})$.
\end{enumerate}
\end{definition}

\begin{figure}[ht]
\includegraphics[width=9cm]{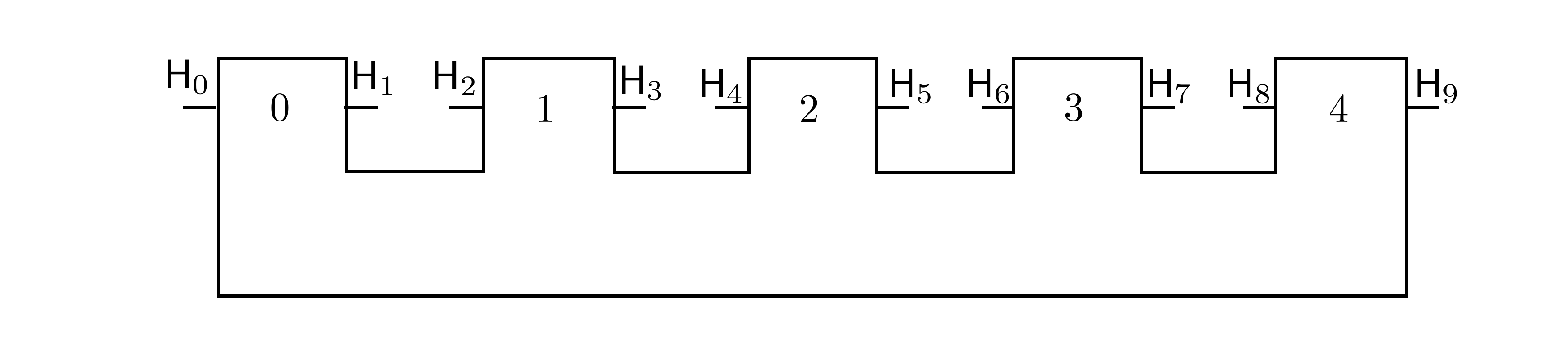}
\caption{Graphical representation of a 5-comb on the spaces $\sH_i$ with $0\leq i\leq 9$. The action on a 4-comb on spaces $\sH_i$ with $1\leq i\leq 8$ is obtained by connecting the identified spaces. Diagrammatically, this corresponds to connecting the identified wires, thus connecting the teeth of the input comb in the slots of the map comb. In formula, this corresponds to the usual representation of the action of a linear map in the Choi representation: $\tT(R_{\mathrm{in}})=\Tr_{12345678}[(I_{09}\otimes R_{\mathrm{in}}^T)\ch(\tT)]$.}
\label{f:combdiag}
\end{figure}

The pair of spaces $\sH_{2j,2j+1}$ identifies the $j+1$-th {\em tooth} of a comb, where the nomenclature is due to the graphical representation of combs as in fig.~\ref{f:combdiag} (see Refs.~\cite{supermaps,comblong})

The main theorems in the theory of combs are the following
\begin{theorem}
A positive operator $R$ on $\sH_{\rA_{2n-1}}\otimes\sH_{\rA_{2n-2}}\otimes\ldots\otimes\sH_{\rA_0}$ belongs to $\det(\rn_{01\ldots(2n-1)})$ iff it satisfies the following constraint
\begin{align}
\Tr_{\rA_{2j+1}}[R^{(j)}]&=I_{\rA_{2j}}\otimes R^{(j-1)},\quad j\geq1,\nonumber\\
\Tr_{\rA_{1}}[R^{0}]&=I_{\rA_0},\nonumber\\
R^{(n-1)}&:=R.
\label{eq:cascatr}
\end{align}
\label{th:cascatr}
\end{theorem}

\begin{theorem}
A positive operator $R$ on $\sH_{\rA_{2n-1}}\otimes\sH_{\rA_{2n-2}}\otimes\ldots\otimes\sH_{\rA_0}$ belongs to $\det(\rn_{01\ldots(2n-1)})$ iff it is the Choi operator of a channel with memory from the ordered input systems $\rA_0\rA_2\ldots\rA_{2n-2}$ to the output ones $\rA_1\rA_3\ldots\rA_{2n-1}$.\label{th:combreal}
\end{theorem}

Theorem \ref{th:combreal} asserts that a comb in $\det(\rn_{01\ldots(2n-1)})$ can be realised by a circuit as follows
\begin{align}
\begin{aligned}
    \Qcircuit @C=1em @R=.7em @! R {&&\pureghost{\tA_1}&\ustick{\sH_1}\qw&\qw&&&&&&&&\\
    &\ustick{\sH_0}\qw&\multigate{-1}{\tA_1}&\ustick{\sH_{A_1}}\qw&\multigate{1}{\tA_2}&\ustick{\sH_3}\qw&\qw&&&&&&\\
     &&&\ustick{\sH_2}\qw&\ghost{\tA_2}&\ustick{\sH_{A_2}}\qw&\qw&&&&&&\\
     &&&&&&\ddots&&&&&&\\
     &&&&&&&\qw&\ustick{\sH_{A_{n-1}}}\qw&\qw&\multigate{1}{\tA_n}&\qw&\ustick{\sH_{2n-1}}\qw&\qw\\
     &&&&&&&\qw&\ustick{\sH_{2n-2}}\qw&\qw&\ghost{\tA_{n}}&&
     }
\end{aligned}
%\begin{aligned}
%    \Qcircuit @C=1em @R=.7em @! R {&\ustick{\sH_0}\qw&\multigate{1}{\tA_1}&\ustick{\sH_1}\qw&\qw&&\ustick{\sH_2}\qw&\multigate{1}{\tA_2}&\ustick{\sH_3}\qw&\qw&&\ldots&&\ustick{\sH_{2n-2}}\qw&\qw&\multigate{1}{\tA_n}&\qw&\ustick{\sH_{2n-1}}\qw&\qw\\
%    &&\pureghost{\tA_1}&\qw&\qw&\ustick{\sH_{A_1}}\qw&\qw&\ghost{\tA_2} &\qw&\ustick{\sH_{A_2}}\qw&\qw&\ldots&\qw&\ustick{\sH_{A_{n-1}}}\qw&\qw&\ghost{\tA_{n}}&&}
%\end{aligned}
\end{align}

In the following we will prove characterisation theorems that depend on the depth of combs, summarised by the integer $\rn$, and are independent of the particular dimension of the spaces $\sH_0,\sH_1,\ldots,\sH_{2n-1}$. For this reason, we will often refer to the general class of $\rn$-combs by the type $\rn$, dropping the labels of spaces. Moreover, it will be useful to consider classes of $\rn$-combs on the same spaces, but with permuted teeth. For this reason, we will introduce the notation $\rn_\sigma$, meaning that for any given space $\sH_{2n-1}\otimes\sH_{2n-2}\otimes\ldots\otimes\sH_{0}$, $\rn_\sigma$ encompasses the type of $\rn$-combs on $\sH_{2\sigma(n-1)+1}\otimes\sH_{2\sigma(n-1)}\otimes\ldots\otimes\sH_{2\sigma(0)}$.

\subsection{Maps from combs to combs}

The next step is to prove a characterisation theorem for maps from combs to combs. For this purpose, it is useful to prove some preliminary lemmas, providing a clearer picture of the structure of the maps. In particular, the results presented in this section are useful in identifying the general structure of spaces $\Delta_{\mathrm m\to \mathrm n}$ that only depend on the numbers $\mathrm m$ and $\mathrm n$ of teeth, and not on the dimensions $d_{\rA_i}$ of the involved systems $\rA_i$.

The first result that we need is a characterisation fo the space $\Delta_\rem$ in terms of spaces $\sL_\vb$.
\begin{lemma}
The space $\Delta_\rem$ is the direct sum
\begin{align}
\bigoplus_{\vb\in\set E_1}\sL_\vb,
\end{align}
where $\set E_1$ is the set of binary strings start with an even number of $1$'s and that have at least one 0.
\end{lemma}
\Proof This characterisation immediately follows from theorem \ref{th:cascatr}.\qed

\begin{corollary}
Let $p=m+n$. Then one has
\begin{align}
\Delta_{\rp}&=(\Gamma_{\rem}\otimes\Delta_\rn)\oplus(\Delta_\rem\otimes\Gamma_\rn)\nonumber\\
&\oplus(\Delta_\rem\otimes\Delta_\rn)\oplus(\Delta_\rem\otimes\overline\Delta_\rn).
\label{eq:deltampiun}
\end{align}
\label{cor:deltampiun}
\end{corollary}

Let us now consider the types $\rem\otimes \rn$. By equation \eqref{eq:afftens} the general element of $\det(\rem\otimes\rn)$ is an affine combination of tensor products $M\otimes N$, with $M\in\det(\rem)$ and $N\in\det(\rn)$. Considering each term of the affine combination separately, it is easy to check that if we arrange the $m$ teeth of the first comb to the left and the $n$ teeth of the second to the right, elements of $\det(\rem\otimes\rn)$ satisfy condition \ref{eq:cascatr}, and thus  belong to $\det(\rp)$ with $p=m+n$. Moreover, the same result holds if we permute the teeth of the $\rp$-comb in such a way that the ordering of teeth of the $\rem$-comb and that of teeth of the $\rn$-comb are preserved. We denote the set of these permutations as $\Sigma_{\rem,\rn}$. For example, let $\rem=\rn=2$. In this case we have two combs, both having two teeth. Let us label the teeth of the first comb by $0,1$ and those of the second by $2,3$. The starting arrangement is thus $0,1,2,3$. The allowed permutations are all the permutations that do not bring the tooth $1$ to the left of $0$ or $3$ to the left of $2$, namely
\begin{align*}
&0,1,2,3,\\
&0,2,1,3,\\
&0,2,3,1,\\
&2,0,1,3,\\
&2,0,3,1,\\
&2,3,0,1,
\end{align*}
that is
\begin{align}
\Sigma_{2,2}=\{(12),(123),(021),(0231),(02)(13)\}.
\label{eq:perms}
\end{align} 
We now formalise the above argument by the following statement.
\begin{lemma}
The space $\Delta_{\rem\otimes\rn}$ is contained in the intersection of the spaces $\Delta_{(\rem+\rn)_\sigma}$, where $\sigma\in\Sigma_{\rem,\rn}$. In formula,
\begin{align}
\Delta_{\rem\otimes\rn}\subseteq\bigcap_{\sigma\in\Sigma_{\rem,\rn}}\Delta_{\rp_\sigma}
\end{align}
\label{lem:orderintersec}
\end{lemma}
%\Proof Let $R$ be an element of $\rem\otimes\rn$. $\sigma\in\Sigma_{\rem,\rn}$. If we permute the $m+n$ teeth by the permutation $\sigma$, by construction we have
We can now evaluate the cardinality of $\Sigma_{\rem,\rn}$ through the following lemma.

\begin{lemma}
Let $\rm$ and $\rn$ be two comb types. The cardinality of $\Sigma_{\rem,\rn}$ is
\begin{align}
|\Sigma_{\rem,\rn}|={{m+n}\choose{n}}
\end{align}
\end{lemma}
\Proof The proof proceeds as follows. Let us consider an ordered array of $n+m$ slots, in which we will allocate the teeth of the two combs. Every different allocation results in a different permutation. For example, for $m=2$ and $n=3$ the array has length 5, and one has the following possible allocations
\begin{align*}
&mmnnn,\ mnmnn,\ mnnmn,\ mnnnm,\ nmmnn,\\
&nmnmn,\ nmmnn,\ nnmmn,\ nnmnm,\ nnnmm.
\end{align*}
In the general case, we can think of an allocation as a choice of a subset of $m$ slots out of the total $m+n$ slots. The number of subsets with $m$ elements of a set of $m+n$ elements is precisely the number of combinations of $m+n$ elements of class $m$, whose cardinality is well known to be ${m+n\choose m}$.\qed

The last permutation in equation \eqref{eq:perms} completely reverses the order of the two combs. In the general case, the permutation that exchanges the two combs---denoted in the following by $\sigma_\leftrightarrow$---always belongs to $\Sigma_{\rem,\rn}$, and plays a special role in the next results.

\begin{lemma}
Let $\rem$ and $\rn$ be two comb types, and let $\rp$ be the comb type with $p=m+n$ corresponding to the arrangement of the teeth of the $\rem$-comb to the left and those of the $\rn$-comb to the right. Then we have
\begin{align}
\Delta_{\rp\cap\rp_{\sigma_\leftrightarrow}}=\Delta_{\rem\otimes\rn}.
\end{align}
\label{lem:fiko}
\end{lemma}
\Proof By equation \eqref{eq:deltampiun} one has
\begin{align*}
\Delta_{\rp\cap\rp_{\sigma_\leftrightarrow}}=(\Gamma_{\rem}\otimes\Delta_\rn)\oplus(\Delta_\rem\otimes\Gamma_\rn)\nonumber\oplus(\Delta_\rem\otimes\Delta_\rn),
\end{align*}
and by equation \eqref{eq:deltaxoty} the thesis follows.\qed

Finally, we can now prove the following crucial result.
\begin{theorem}
Let $\rem$ and $\rn$ be two comb types. Then one has
\begin{align}
\Delta_{\rem\otimes\rn}=\bigcap_{\sigma\in\Sigma_{\rem,\rn}}\Delta_{(\rem+\rn)_\sigma}.
\end{align}
\label{th:interst}
\end{theorem}
\Proof By lemma \ref{lem:orderintersec} one has \begin{align*}
\Delta_{\rem\otimes\rn}\subseteq\bigcap_{\sigma\in\Sigma_{\rem,\rn}}\Delta_{\rp_{\sigma}}\subseteq\Delta_{\rp\cap\rp_{\sigma_\leftrightarrow}}=\Delta_{\rem\otimes\rn}.
\end{align*} 
Thus, the two inclusions are actually equalities, and we have
\begin{align*}
\Delta_{\rem\otimes\rn}=\bigcap_{\sigma\in\Sigma_{\rem,\rn}}\Delta_{\rp_{\sigma}}=\Delta_{\rp\cap\rp_{\sigma_\leftrightarrow}}.
\end{align*} 
\qed

We now use the above theorem to prove the main result in this section, which provides a characterisation of maps from $\rem$-combs to $\rn$-combs.

\begin{theorem}
For maps of type $\rem\to \rn$ one has
\begin{align}
\Delta_{\rem\to\rn}=\Span(\Delta_{(\rem+\rn-1)\to1}\cup\Delta_{(\rem+\rn-1)_{\sigma_\leftrightarrow}\to1}),
\end{align}
and
\begin{align}
\det&({\rem\to\rn})=\con(\rem\to\rn)\nonumber\\
&\cap\aff(\det[(\rem+\rn-1)\to1]\cup\det[(\rem+\rn-1)_{\sigma_\leftrightarrow}\to1]),
\end{align}
where $\sigma_{\leftrightarrow}$ is the permutation that exchanges the $\rem$-comb with the $\rn-1$-comb representing the input type of the output $\rn$-comb.
\label{th:tombstone}
\end{theorem}
\Proof First of all, we remind that $\rn=(\rn-1)\to1$, and by Corollary \ref{cor:uncurtyp}, 
\begin{align*}
\rem\to\rn=&\rem\to[(\rn-1)\to1]\\
=&[\rem\otimes(\rn-1)]\to1.
\end{align*}
Now, thanks to Lemma \ref{lem:fiko} we have 
\begin{align*}
\rem\otimes(\rn-1)=(\rem+\rn-1)\cap(\rem+\rn-1)_{\sigma_{\leftrightarrow}},
\end{align*}
and finally by Lemma \ref{lem:deltacapto}
\begin{align*}
\Delta_{\rem\to\rn}=\Span(\Delta_{(\rem+\rn-1)\to1}\cup\Delta_{(\rem+\rn-1)_{\sigma_\leftrightarrow}\to1}).
\end{align*}
We can also use Corollary \ref{cor:detaffcap} to conclude that
\begin{align*}
&\det({\rem\to\rn})=\con(\rem\to\rn)\nonumber\\
&\ \cap\aff\{\det[(\rem+\rn-1)\to1]\cup\det[(\rem+\rn-1)_{\sigma_\leftrightarrow}\to1]\}.
\end{align*}
\qed
%is a special type of $\rn+1$-comb (see lemma \ref{lem:*}). Moreover, $\rem\to\rn=\overline{\rem\otimes\overline\rn}$. Now, by Theorem \ref{th:interst} we have 
%\begin{align*}
%\Delta_{\rem\otimes \overline\rn}=\bigcap_{\sigma\in\Sigma_{\rem,\overline\rn}}\Delta_{(\rem+\rn+1)_\sigma}.
%\end{align*}
%Finally, by Lemma \ref{lem:complemint} we have
%\begin{align*}
%\Delta_{\rem\to\rn}=\Delta_{\overline{\rem\otimes \overline\rn}}&=\overline{\bigcap_{\sigma\in\Sigma_{\rem,\overline\rn}}\Delta_{(\rem+\rn+1)_\sigma}}\\
%&=\Span\left(\bigcup_{\sigma\in\Sigma_{\rem,\overline\rn}}\Delta_{\overline{(\rem+\rn+1)_\sigma}}\right).
%\end{align*}
%\qed

\begin{figure}[ht]
\includegraphics[width=8.1cm]{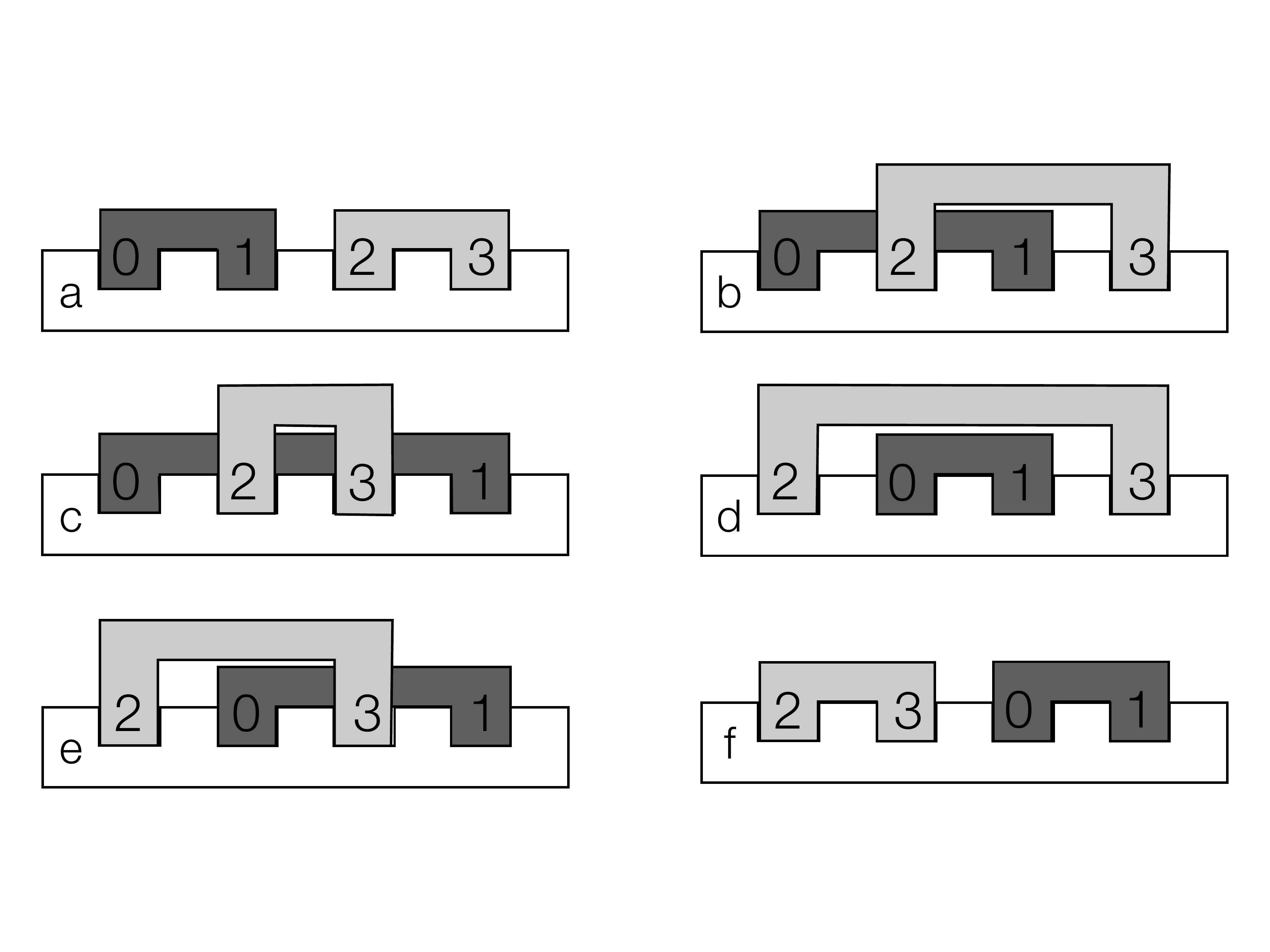}
\caption{Graphical representation of the structure of a map from 2- to 3-combs. The possible orderings of the $2\to 3$ map correspond to the six white combs on the bottom. The input 2-comb is the one represented in dark grey, with teeth labelled $0,1$. The output 3-comb is a map acting on the light grey 2-comb with teeth labelled $2,3$. According to theorem \ref{th:tombstone}, the only the structures that are necessary to define a map $2\to3$ are those of of diagrams (a) and (f).\label{fig:affcomb}}
\end{figure}

Thanks to Lemma \ref{lem:compcapdet}, we can figure out the meaning of the above theorem as follows. The most general maps from $\rem$-combs to $\rn$-combs are represented by affine combinations of $\rem+\rn+1$-combs with orderings given by those permutations $\sigma$ of teeth that are compatible with both the teeth ordering of input $\rem$-combs and of output $\rn$-combs. A more intuitive picture of the general map $\rem\to\rn$ is provided in Fig.~\ref{fig:affcomb} for the case $m=2$, $n=3$. 

%*****************
%
%\begin{theorem}
%Every deterministic event of type $x$ is an affine combination of tensor products of combs, each of which has either one of a set of orderings, or simultaneously all the orderings in a set.
%\end{theorem}
%\Proof Once again, the proof proceeds by induction. Suppose that the thesis is true for $x$, i.e.~$x=\bigcup_i(x^{(i)}_1\cap x^{(i)}_2)$. Then $\overline x=\bigcap_i(\overline x^{(i)}_1\cup\overline x^{(i)}_2)$ 
%
%\begin{align*}
%&(A_{11}\cup A_{12})\cap(A_{21}\cup A_{22})\\
%=&(A_{11}\cap A_{21})\cup(A_{11}\cap A_{22})\cup(A_{12}\cap A_{21})\cup(A_{12}\cap A_{22})\\
%&\bigcap_{j=1}^M\bigcup_{k=1}^N A_{j,k}=\bigcup_{k\in\{1,\ldots,N\}^M}\bigcap_{j=1}^M A_{j,k_j}.
%\end{align*}

\section{Conclusion}\label{s:conc}

We reviewed the main points of the theory of combs, i.e.~maps from quantum circuits into quantum channels (or more generally quantum operations), reporting the crucial realisation theorem, which asserts that combs are physically obtained by circuits with open slots. We then focused our attention on the hierarchy of all mathematical maps, from combs to combs and maps thereof, that are admissible, that is to say consistent with the properties of probabilities. We introduced a language of types and appropriate typing rules, with a partial ordering of types that allows for proofs by induction, and used induction to prove general structure theorems for the set of admissible maps of any type. In particular, we showed that maps at every order in the hierarchy  inherit normalisation constraints from the first-level causality constraints. However, most of higher-order maps require indefinite causal structures for their implementation. We then restricted attention to maps from combs to combs. We first showed that such maps can be seen as maps from tensor products of combs into channels. We then characterised them as those maps that can be represented as affine combinations quantum combs with two different orderings, the first one treating the input tensor product $A\otimes B$ as a comb where the teeth of $A$ precede those of $B$, and the other one treating $A\otimes B$ as a comb where the teeth of $B$ precede those of $A$. This result provides a great simplification of the general structure of maps from combs to combs.

The surprising issue with the hierarchy of higher order quantum maps is that, while for quantum combs the admissibility constraint are necessary and sufficient for the existence of an implementation scheme, in the case of higher-order maps such equivalence seems to be beyond our present understanding of physics, and possibly requires a theory that encompasses quantum information theory and a theory of indefinite causal orderings, such as general relativity. The problem of implementation thus remains open, leaving three different possibilities: i) all admissible maps are achievable in a futuristic quantum-gravity scenario; ii) there is some polynomially computable constraint beyond admissibility that separates feasible from unfeasible maps; iii) the distinction is given by a non-computable constraint, which essentially means that, given the Choi representation of an admissible higher-order map, it is not possible to say a priori whether it represents a feasible computation, and the answer can be given only in some special case. The last situation represents to some extent a generalisation of the problem of determining whether a given density matrix describes a quantum state that is entangled or separable.

%The construction of types on Sect. \ref{s:hier} corresponds in abstract terms to the rules of a $*$-autonomous category, whose type theory is a model of Multiplicative Linear Logic (MLL). Decidability of implication in MLL corresponds to decidability of type inclusions in a $*$-autonomous category \footnote{A. Kissinger and S. Uijlen, private communication}.

\acknowledgments The author is grateful to Aleks Kissinger and Fabio Costa for useful discussions.

\bibliography{biblio}
%\begin{thebibliography}{99}
%\end{thebibliography}
\end{document}